\documentclass[12pt]{article}

\usepackage{epsfig}
\usepackage{amsmath}
\usepackage{amssymb}
\usepackage{axodraw}

\begin{document}

\begin{titlepage}
\begin{flushright}
\today
\end{flushright}

\vspace{1cm}
\begin{center}
\baselineskip25pt {\Large\bf Dual Brane Pairs, Chains and the
Bekenstein-Hawking Entropy}

\end{center}
\vspace{1cm}
\begin{center}
\baselineskip12pt {Axel Krause\footnote{E-mail: {\tt
krause@schwinger.harvard.edu}, now at Jefferson Physical
Laboratory, Harvard University, Cambridge, MA 02138, USA}}
\vspace{1cm}

{\it Physics Department}\\[1.8mm]
{\it National Technical University of Athens}\\[1.8mm]
{\it 15773 Athens, Greece}

\vspace{0.3cm}
\end{center}
\vspace*{\fill}

\begin{abstract}
A proposal towards a microscopic understanding of the
Bekenstein-Hawking entropy for D=4 spacetimes with event horizon
is made. Since we will not rely on supersymmetry these spacetimes
need not be supersymmetric. Euclidean D-branes which wrap the
event horizon's boundary will play an important role. After
arguing for a discretization of the Euclidean D-brane worldvolume
based on the worldvolume uncertainty relation, we count chainlike
excitations on the worldvolume of specific dual Euclidean brane
pairs. Without the need for supersymmetry it is shown that one can
thus reproduce the D=4 Bekenstein-Hawking entropy and its
logarithmic correction.
\end{abstract}

\vspace*{\fill}

\end{titlepage}

\section{Introduction}

Three decades ago it was proposed by Bekenstein \cite{Bek} to
associate a physical entropy with a black hole. This entropy was
argued to be proportional to the area $A_H$ of the black hole's
horizon. Evidence for this proposal came from earlier work of
Christodoulou \cite{Chr} and Hawking \cite{Haw1}. Christodoulou
had shown that for physical processes which result in the
absorption of a particle by a Kerr black hole its so-called
irreducible mass cannot decrease but only increase. The
irreducible mass is proportional to $\sqrt{A_H}$. Hawking, then
followed with a general proof that $A_H$ cannot decrease in any
classical physical process.

For a consistent description of black holes by thermodynamic
quantities it is necessary that they emit thermal radiation at a
temperature compatible with the laws of thermodynamics. This was
indeed found by Hawking \cite{Haw2} and required the inclusion of
quantum effects. With the derived value for the
Hawking-temperature $T_H$ it was then possible to fix the
proportionality constant in the entropy-area relation and to
assign to each black hole the Bekenstein-Hawking (BH) entropy
\begin{equation}
{\cal S}_{BH} = \frac{A_H}{4G_4}\frac{k_Bc^3}{\hbar} \; .
\label{SBH}
\end{equation}
The appearance of Planck's Constant already points to the fact --
as suggested by Statistical Mechanics -- that one should better
understand and count the microscopic quantum mechanical degrees of
freedom leading to the formation of the black hole in order to
understand the BH-entropy at a fundamental level.

Later it was shown by Gibbons and Hawking \cite{GH} that
(\ref{SBH}) also applies to cosmological de Sitter event horizons
to which one can similarly ascribe a thermal Hawking-temperature.
Thus black hole and cosmological event horizons should share a
common underlying microscopic property giving rise to the same
universal expression for their entropy. It is the aim of this
paper to propose a set of microscopic states which can account for
the BH-entropy and its logarithmic corrections in a rather
universal way. We will lay out here the general framework and
reserve the particular application to Schwarzschild black holes
and de Sitter cosmologies to future work.

Both string-theory and the quantum geometry program made decisive
steps in recent years towards an identification of the microscopic
black hole states. While efforts in string-theory, which increased
dramatically after the work of \cite{SV}, more or less focussed on
supersymmetric black holes or small deviations thereof, the
Quantum Geometry approach led to the derivation of the BH-entropy
for the D=4 Schwarzschild case \cite{ABCK} but had to fix an
undetermined multiplicative factor, the Barbero-Immirzi parameter
which arises from an ambiguity in the loop quantisation procedure,
appropriately. Starting from string- resp.~M-theory we want to
take here a different route trying to address the problem of
understanding the BH-entropy directly. One of the interesting
features of our proposal will be that although starting from
branes in string-theory we are led to chain-like excitations on
the branes' worldvolume which resemble the polymer-like
excitations of Quantum Geometry (see e.g.~\cite{ARev}). This
raises at least the hope that there might be some reconciliation
between these two major approaches to quantum gravity. We will
focus in this work on the case of D=4 spacetimes with event
horizons. The generalization to higher dimensions is presented in
\cite{KHigher} while further aspects of black holes are addressed
in \cite{KBlack}.

\section{Dual Brane Pairs and the BH-Entropy}

{\sl String-Theory Case:} Consider type II string-theory on a D=10
spacetime with Lorentz\-ian signature which factorizes into ${\cal
M}^{1,3}\times{\cal M}^{p-1}\times{\cal M}^{7-p}$ ($p=1,\hdots,5$)
and is described by a metric
\begin{equation}
ds^2 = g_{\mu\nu}^{(1,3)}(x^\rho)dx^\mu dx^\nu
+ g_{ab}^{(p-1)}(x^c)dx^adx^b + g_{kl}^{(7-p)}(x^m)dx^kdx^l
\; , \label{Metric}
\end{equation}
where
\begin{equation}
\mu,\nu=0,\hdots,3\, ; \qquad a,b,c=1,\hdots,p-1\, ; \qquad
k,l,m=1,\hdots,7-p \; . \notag
\end{equation}
The metric $g_{\mu\nu}$ describes a D=4 spacetime geometry of
which we assume that it possesses an event horizon with associated
BH-entropy. The 2-surface $H$ will represent the horizon's
boundary (not to be confused with ${\cal H}^+$, the future event
horizon, or ${\cal H}^-$ the past event horizon; in the case of a
D=4 Schwarzschild black hole the boundary $H$ is a 2-sphere
$S_H^2$ defined as the intersection of the future event horizon
${\cal H}^+$ with a partial Cauchy surface ending at spatial
infinity ${\cal I}^0$ in the exterior black hole spacetime).
Moreover, to describe a compactification from D=10 down to D=4 we
take the two internal ${\cal M}^{i}$ to be compact. A special
example with constant internal metric would be a
$T^6=T^{p-1}\times T^{7-p}$ torus-compactification. For these
backgrounds the effective D=4 Newton's Constant is related to the
Regge slope $\alpha'$ and the string coupling constant $g_s$
through (we employ conventions as given in \cite{CVJ})
\begin{equation}
G_4=\frac{G_{10}}{V_{p-1}V_{7-p}}=\frac{(2\pi)^6{\alpha'}^4g_s^2}{8V_{p-1}
V_{7-p}} \; ,
\end{equation}
where $V_i=vol({\cal M}^{i})\equiv\int_{{\cal M}^{i}}d^ix
\sqrt{g^{(i)}}$.

Imagine now wrapping two orthogonal Euclidean `electric-magnetic'
dual branes, $Dp$ and $D(6-p)$, around $H\times{\cal M}^{p-1}$ and
${\cal M}^{7-p}$, respectively. So, together the two branes cover
the whole internal space plus the area of the exterior D=4
spacetime's boundary. For such a dual brane pair it follows from
the Dirac-quantisation condition \cite{DNT} that the product of
their tensions is given by
\begin{equation}
\tau_{Dp}\tau_{D(6-p)}=\frac{1}{(2\pi)^6{\alpha'}^4g_s^2} \; .
\end{equation}
Thus we can write
\begin{equation}
\frac{1}{G_4}=8(\tau_{Dp}V_{p-1})(\tau_{D(6-p)}V_{7-p}) \; .
\end{equation}
Due to the fact that part of the $Dp$ brane wraps the full area of
the D=4 spacetime's boundary we may rewrite the BH-entropy of the
D=4 spacetime as
\begin{equation}
{\cal S}_{BH}=\frac{A_H}{4G_4}=2S_{Dp}S_{D(6-p)} \; ,
\label{Rexpress}
\end{equation}
where
\begin{equation}
S_{Dp}=\tau_{Dp}\int_{H\times{\cal M}^{p-1}} d^{p+1}x\sqrt{\det g}
\; , \qquad S_{D(6-p)}=\tau_{D(6-p)}\int_{{\cal M}^{7-p}}
d^{7-p}x\sqrt{\det g}
\end{equation}
are the respective Nambu-Goto actions of the dual branes. For our
later analysis it turns out to be necessary to get rid of the
factor two on the rhs of (\ref{Rexpress}). This can easily be
achieved by considering another Euclidean brane pair.

Let us therefore take a second dual Euclidean brane pair $Dq-
D(6-q)$, ($q=1,\hdots,5$) and consider a D=10 spacetime background
of the form ${\cal M}^{1,3}\times{\cal M}^{p-1}\times {\cal
M}^{q-p}\times{\cal M}^{7-q}$ (without loss of generality we can
assume that $p\le q$) with internal metric
\begin{equation}
g_{ab}^{(p-1)}(x^c)dx^adx^b + g_{qr}^{(q-p)}(x^s)dx^qdx^r +
g_{uv}^{(7-q)}(x^w)dx^udx^v \; .
\label{Metric2}
\end{equation}
We wrap the branes of the first pair $Dp$, $D(6-p)$ around
$H\times{\cal M}^{p-1}$ resp.~${\cal M}^{q-p}\times{\cal M}^{7-q}$
while the branes of the second pair, $Dq$ and $D(6-q)$ wrap
$H\times{\cal M}^{p-1}\times{\cal M}^{q-p}$ resp.~${\cal
M}^{7-q}$. By repeating the steps which led to (\ref{Rexpress})
one may now express the D=4 BH-entropy as
\begin{equation}
{\cal S}_{BH}=S_{Dp}S_{D(6-p)}+S_{Dq}S_{D(6-q)} \; .
\label{SBH2}
\end{equation}
Notice further that we are free to exchange any of the appearing
D-branes by its anti-D-brane and still arrive at the same
expression (\ref{SBH2}). This option becomes important when one
wants to address uncharged non-supersymmetric D=4 spacetimes. We
can therefore finally state that for all such dual Euclidean brane
configurations the D=4 BH-entropy can be universally,
i.e.~irrespective of the particular choice of dual brane pairs,
rewritten as
\begin{equation}
{\cal S}_{BH}=\sum_{i=1,2}S_{E_i}S_{M_i} \; ,
\label{SBH3}
\end{equation}
where
\begin{equation}
(E_i,M_i) \in \{(Dp_i,D(6-p_i)),(\overline{D}p_i,D(6-p_i)),
(Dp_i,\overline{D}(6-p_i)),(\overline{D}p_i,\overline{D}(6-p_i)\}
\label{ChoiceD}
\end{equation}
and $S_{E_i}$, $S_{M_i}$ are the respective Nambu-Goto actions for
these branes. The connection between a specific pair of branes and
a specific D=4 spacetime should be established on the basis of the
D=10 spacetime which the pair of branes creates as gravitational
sources followed by a dimensional reduction to four dimensions.

Let us make two comments. First, until now we have restricted the
range of the D-brane dimensions to $p_i=1,\hdots,5$. The reason
being that a Euclidean D0 brane cannot cover the whole $H$.
Therefore, if e.g.~$H=S^2$ the metric of the sphere does not
factorize into two independent 1-dimensional parts and one
therefore cannot write a product of two Nambu-Goto actions in this
situation. However, there is no problem with the case where a $D6$
or $\overline{D}6$ wraps $H\times{\cal M}^5$ and the dual $D0$ or
$\overline{D}0$ the remaining internal ${\cal M}^1=S^1$. With this
subtlety in mind we can extend the range to $p_i=0,\hdots,6$.
Second, since $\tau_{F1}\tau_{NS5}=\tau_{Dp}\tau_{D(6-p)}$ we
could also extend our treatment to incorporate Euclidean
$(F1,NS5)$ as dual pairs. It is easy to see that wrapping a
Euclidean fundamental string $F1$ over $H$ (resp.~the internal
${\cal M}^2$) and the dual Euclidean $NS5$ on the complete
internal space ${\cal M}^6$ (resp.~$H\times {\cal M}^4$) in
combination with a second $D1$-$D5$ pair or another $(F1,NS5)$
pair leaves (\ref{SBH3}) intact. Hence we can enlarge the set of
dual pairs (\ref{ChoiceD}) to include also
\begin{equation}
(F1,NS5) \; .
\end{equation}
Moreover both $F1$ and $NS5$ can be replaced independently by
their NS-NS charge reversed antipartners.

{\sl M-Theory Case:} Since (\ref{SBH3}) works so universally for
all dual pairs of type II string-theory one should expect the
formula also to hold true for the unique M-theory dual brane pair,
the $(M2,M5)$ pair. This is what we will show now. Let us start
again with a single Euclidean $(M2,M5)$ pair where the $M2$ wraps
the boundary $H$ associated with the D=4 spacetime's horizon plus
an internal $S^1$ while the Euclidean $M5$ wraps the remaining
internal six-space ${\cal M}_6$. That means we assume the metric
of the D=11 spacetime to factorize into the direct product
structure ${\cal M}^{1,3}\times S^1\times{\cal M}^6$.

The tensions of the dual branes satisfy ($l_{Pl}$ is the D=11
Planck-length)
\begin{equation}
\tau_{M2}\tau_{M5}=\frac{1}{(2\pi)^7l_{Pl}^9}
\end{equation}
and $G_4$ can be expressed in terms of the D=11 Newton's constant
as
\begin{equation}
G_4=\frac{G_{11}}{LV_6}=\frac{(2\pi)^7l_{Pl}^9}{8LV_6}
\end{equation}
where $L=vol(S^1),V_6=vol({\cal M}^6)$. So combining these two
equations we obtain
\begin{equation}
\frac{1}{G_4}=8(\tau_{M2}L)(\tau_{M5}V_6) \; .
\end{equation}
Because the Euclidean $M2$ partly wraps the full boundary area
$A_H=vol(H)$ we can rewrite the D=4 BH-entropy again in terms of
the Euclidean Nambu-Goto actions of the respective branes
\begin{equation}
{\cal S}_{BH}=\frac{A_H}{4G_4}=2S_{M2}S_{M5} \; .
\end{equation}
In the same manner as for the D-brane case in D=10 we will now add
a second dual Euclidean pair $(M2,M5)$ wrapping likewise the
complete $H \times S^1 \times {\cal M}^6$. This once again allows
us to rewrite the D=4 BH-entropy exclusively in terms of the
Nambu-Goto actions of the involved M-branes
\begin{equation}
{\cal S}_{BH} = \sum_{i=1,2} S_{M2_i} S_{M5_i}
\label{SBHM}
\end{equation}
eliminating the prefactor two. Instead of wrapping each $M2$
around $H$ we could also consider wrapping one or both of the
$M5$'s around $H$ instead. This will lead to internal geometries
which factorize like
\begin{equation}
S^1\times{\cal M}_{(1)}^3\times{\cal M}_{(2)}^3\; ,
\quad {\cal M}^4\times{\cal M}^3
\end{equation}
instead of $S^1\times{\cal M}^6$. It is easy to see that following
the same reasoning as before we end up again with (\ref{SBHM}) in
these situations. Finally replacing any brane by its anti-brane
doesn't change (\ref{SBHM}) because the Nambu-Goto actions stay
invariant under this replacement. So we can conclude that also in
M-theory (\ref{SBH3}) holds true, this time for
\begin{equation}
(E_i,M_i)\in \{ (M2,M5),(\overline{M}2,M5), (M2,\overline{M}5),
(\overline{M}2,\overline{M}5) \} \; .
\end{equation}

\section{Chain-States and Counting of States}

So far we have achieved a universal rewriting of the D=4
BH-entropy in terms of dual pair doublets of type II string-theory
and M-theory. Let us now see in which way the introduction of the
Euclidean pairs can help us to understand the BH-entropy at a
microscopic level by counting an appropriate set of states related
to these pairs.

To set the stage for a proposal of what the microscopic states in
the strongly-coupled regime might be, let us briefly reflect upon
the tension of a brane. Usually a brane's tension $\tau_{Dp}$ (we
take a D-brane for definiteness but the following considerations
apply as well to the $\overline{D}p$ plus the Euclidean
$F1,NS5,M2,M5$ and their antipartners) is conceived as the brane's
mass per unit $p$-volume. This point of view is natural if the
brane's worldvolume has Lorentzian signature and stresses the
split into one time and $p$ space dimensions. However, when
dealing with Euclidean branes it is more natural to treat all
$p+1$ space dimensions on an equal footing. To account for this
let us write the brane's tension as a volume $v_{Dp}$ (with
corresponding length-scale $l_{Dp}$)
\begin{equation}
\tau_{Dp}\equiv\frac{1}{v_{Dp}}=\frac{1}{l_{Dp}^{p+1}} \; .
\end{equation}

Our basic proposition is that the volume $v_{Dp}$ constitutes a
smallest volume unit within the worldvolume of the Euclidean
Dp-brane. Indeed, it suffices to assume that it is only the
entropy-carrying chains to be introduced shortly which cannot
resolve a worldvolume smaller than $v_{Dp}$. This is analogous to
the statement that the weakly-coupled fundamental string cannot
resolve length-scales shorter than the string-scale
$\sqrt{\alpha'}$.

Evidence for such a smallest volume unit on a brane's worldvolume
comes from the `worldvolume uncertainty relation' for D-branes
\cite{CHK} as we will now explain (cf.~also the discussion in
\cite{KRev}). It is well-known that a D-brane in the presence of a
background magnetic flux along its worldvolume acquires a
non-commutative geometry \cite{NG}. The non-trivial commutator
\begin{equation}
[X^i,X^j] = 2\pi i\alpha'{\cal F}^{ij}
\end{equation}
(with ${\cal F}=B-dA$ the difference of the NS-NS 2-form potential
and the $U(1)$ gauge field strength on the brane) of the D-brane's
longitudinal coordinates $X^i$ already suggests a worldvolume
uncertainty relation. Furthermore, it was shown in \cite{CHK} that
a non-trivial expectation value for ${\cal F}$ can also arise from
integrating out quantum fluctuations around a classical
background, even when a background flux ${\cal F}$ is absent. The
proper framework to describe this result is string field theory,
where one describes the D-brane through a normalized wave-function
$\Psi(X)$ with $X$ the brane's target space coordinates. The
quadratic deviation is then given by the functional integral
\begin{equation}
(\Delta X^i)^2 = \int[DX]\Psi(X)^\dagger (X^i - \bar{X^i})^2
\Psi(X)
\end{equation}
with average value
\begin{equation}
\bar{X^i} = \int[DX]\Psi(X)^\dagger X^i \Psi(X) \; .
\end{equation}
Moreover, one notices that ${\cal F}$ as a background field is
independent of $\Psi$. From the commutator $[X^i,X^j] = 2\pi
i\alpha'{\cal F}^{ij}$ one therefore obtains via the standard
quantum mechanical procedure the relation
\begin{equation}
\Delta X^i \Delta X^j \ge 2\pi \alpha' |{\cal F}^{ij}| \; .
\end{equation}
Let us now come to the uncertainty in $X^i$ which is defined
through
\begin{equation}
\delta X^i = \langle(\Delta X^i)^2\rangle^{1/2} \; ,
\end{equation}
where the expectation value is determined via the string field
path integral
\begin{equation}
\langle (\Delta X^i)^2 \rangle = \frac{1}{Z}\int [D{\cal B}]e^{-S}
(\Delta X^i)^2 \; .
\label{SFPI}
\end{equation}
Here $\cal{B}$ comprises all component fields contained in $\Psi$
which includes the metric and $B$. From the Cauchy-Schwarz
inequality one obtains $(\delta X^i)^2(\delta X^j)^2\ge
|\langle\Delta X^i\Delta X^j\rangle|^2$. Therefore the product of
the uncertainties of two different brane coordinates becomes
lower-bounded by the expectation value for $|{\cal F}^{ij}|$
\begin{equation}
\delta X^i \delta X^j \ge 2\pi \alpha' \langle |{\cal
F}^{ij}|\rangle \; .
\end{equation}

In \cite{CHK} the expectation value $\langle |{\cal
F}^{ij}|\rangle$ has been calculated for the $D1$-brane case by
using for the action $S$ in eq.(\ref{SFPI}) the supergravity
action which is a valid approximation at small string coupling. It
was found that even in the absence of a background ${\cal F}$
field, quantum fluctuations lead to a non-trivial expectation
value and thus an uncertainty relation among the $D1$-brane
worldvolume coordinates. In this case with no explicit background
flux the expectation value becomes a simple expression in terms of
the string coupling constant $g_s$ and $\alpha'$. The result for
other $Dp$-branes plus the M-theory $M2$ and $M5$-branes was then
inferred by string-duality arguments. When one expresses all these
worldvolume uncertainty relations for various branes in terms of
their tensions, as was done in \cite{KRev}, \cite{KSUSY}, it leads
to the following result for any brane with $p+1$ dimensional
worldvolume
\begin{equation}
\delta X^0 \hdots \delta X^p \gtrsim \frac{1}{\tau} \; ,
\end{equation}
valid for all $Dp$-branes and $M2$, $M5$ with $\tau$ the
respective brane's tension. One therefore sees that the smallest
volume allowed by this brane worldvolume uncertainty principle is
indeed given by the inverse of the brane's tension which provides
nice evidence for our assumption.

Equipped with this notion of a smallest world volume unit which we
will term a cell henceforth, we are naturally led to think of the
Euclidean $Dp$-brane as being composed out of $N_{Dp}$ cells,
where $N_{Dp}$ is measured by the brane's Nambu-Goto action
\begin{equation}
N_{Dp}=\tau_{Dp}\int d^{p+1}x\sqrt{\det g} \; .
\label{Measure}
\end{equation}
More precisely, since the Nambu-Goto action is ordinarily assumed
to take smooth continuous values, one should regard the Nambu-Goto
action as an approximation to a more fundamental microscopic
integer-valued function. This latter function would give the
number of cells contained in the brane's worldvolume but will be
well approximated by the smooth Nambu-Goto action when the number
of cells becomes large and the discrete cell structure becomes
quasi-continuous. This large-cell limit is the case we are
interested in here.

So each brane pair $(Dp,D(6-p))$ having mutually orthogonal branes
possesses $N_{Dp}\times N_{D(6-p)}$ cells. Thus altogether the
doublet of dual Euclidean pairs used for the reformulation of the
BH-entropy exhibits a total of
\begin{equation}
N=\sum_{i=1,2}N_{E_i}N_{M_i}
\end{equation}
cells on its combined worldvolume (counting cells on $(E_1,M_1)$
and $(E_2,M_2)$ separately). Therefore, by virtue of
(\ref{Measure}) and its generalization to $F1, NS5, M2,M5$ plus
antipartners we see that (\ref{SBH3}) becomes
\begin{equation}
{\cal S}_{BH}=\sum_{i=1,2}N_{E_i}N_{M_i}\equiv N \; .
\label{SBH4}
\end{equation}

Let us now conceive on the lattice of the combined
`$(E_1,M_1)+(E_2,M_2)$' brane worldvolume an $(N-1)$-chain, i.e.~a
chain composed out of $N-1$ successive links where we allow all
links to start and end on any occurring cell (see
fig.\ref{Chains})
\setcounter{figure}{0}
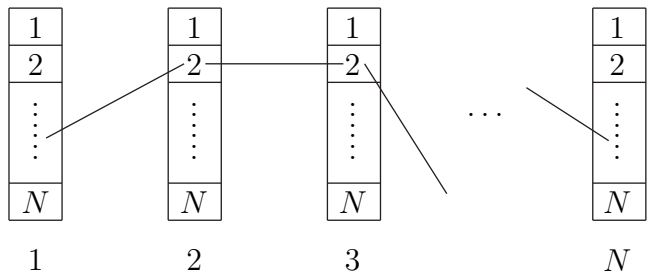
\begin{figure}[t]
\begin{center}
\begin{picture}(260,120)(0,0)
\Text(10,5)[]{$1$}
\Line(0,20)(0,100)
\Line(20,20)(20,100)
\Line(0,100)(20,100)
\Line(0,86)(20,86)
\Line(0,72)(20,72)
\Line(0,34)(20,34)
\Line(0,20)(20,20)
\Text(10,93)[]{$1$}
\Text(10,79)[]{$2$}
\Text(10,63)[]{$\vdots$}
\Text(10,51)[]{$\vdots$}
\Text(10,27)[]{$N$}
\Line(14,51)(66,79)

\Text(70,5)[]{$2$}
\Line(60,20)(60,100)
\Line(80,20)(80,100)
\Line(60,100)(80,100)
\Line(60,86)(80,86)
\Line(60,72)(80,72)
\Line(60,34)(80,34)
\Line(60,20)(80,20)
\Text(70,93)[]{$1$}
\Text(70,79)[]{$2$}
\Text(70,63)[]{$\vdots$}
\Text(70,51)[]{$\vdots$}
\Text(70,27)[]{$N$}
\Line(74,79)(126,79)

\Text(130,5)[]{$3$}
\Line(120,20)(120,100)
\Line(140,20)(140,100)
\Line(120,100)(140,100)
\Line(120,86)(140,86)
\Line(120,72)(140,72)
\Line(120,34)(140,34)
\Line(120,20)(140,20)
\Text(130,93)[]{$1$}
\Text(130,79)[]{$2$}
\Text(130,63)[]{$\vdots$}
\Text(130,51)[]{$\vdots$}
\Text(130,27)[]{$N$}
\Line(134,79)(165,30)

\Text(180,60)[]{$\hdots$}

\Text(230,5)[]{$N$}
\Line(220,20)(220,100)
\Line(240,20)(240,100)
\Line(220,100)(240,100)
\Line(220,86)(240,86)
\Line(220,72)(240,72)
\Line(220,34)(240,34)
\Line(220,20)(240,20)
\Text(230,93)[]{$1$}
\Text(230,79)[]{$2$}
\Text(230,63)[]{$\vdots$}
\Text(230,51)[]{$\vdots$}
\Text(230,27)[]{$N$}
\Line(195,70)(226,50)
\end{picture}
\caption{Constructive view of the $(N-1)$-chain where we arrange
all cells of the lattice in a column and use $N$ copies of them.
We allow each link to connect any cell of a column with any cell
of the succeeding column. Horizontal links correspond to loops.}
\label{Chains}
\end{center}
\end{figure}
In particular a link might start and end on the same cell thus
creating a loop. Altogether the number of such chains is $N^N$.

Our prime motivation to consider long (having about the same
number of links as there are cells in the lattice) chains comes
from the following heuristic reasoning. The `worldvolume
uncertainty principle' for branes led to a smallest resolvable
worldvolume $v_{Dp}=1/\tau_{Dp}$ given by the inverse of the
brane's tension. Hence it implies a minimal resolvable
length-scale $l_{Dp}=v_{Dp}^{1/(p+1)}$. Now, such a minimal
length-scale leads via the ordinary Heisenberg uncertainty
principle of quantum mechanics, involving space coordinates and
their conjugate momenta, to a non-vanishing momentum $\Delta
P\simeq 1/l_{Dp}$. For any relativistic object, for which we can
equate energy with momentum, we are therefore led to a
corresponding energy of magnitude
\begin{equation}
\Delta E \simeq \Delta P \simeq
\frac{1}{l_{Dp}}=(\tau_{Dp})^{\frac{1}{p+1}}
=\frac{1}{\sqrt{\alpha'}(g_s(2\pi)^p)^{\frac{1}{p+1}}}  \; .
\end{equation}
If we associate with $\Delta E$ a temperature, we see that in the
strong-coupling regime, where $g_s\simeq 1$, this temperature is
of the size of the Hagedorn-temperature. Close to this temperature
we know from experience with weakly coupled string-theory that it
is entropically favourable to allocate the energy of the system to
just one single long string instead of distributing the energy
more democratically in smaller portions to lots of small strings
(for a review see \cite{AA}; the relation between entropy and the
length of a string is discussed in \cite{HRS}). This motivates us
to consider as candidates for microscopic excitations chains which
are long\footnote{While long chains are important for
gravitational aspects, short chains composed of just two links
showed up in standard model like constructions \cite{CC1} based on
warped backgrounds \cite{CC2}.}.

The chain-counting until now assumed that all cells were
distinguishable. This however is likely to be changed in a quantum
treatment of the problem. Here the cells would have to be regarded
as `partons', i.e.~indistinguishable bosonic degrees of freedom.
The cure for this is well-known from statistical mechanics. We
have to divide the classical number of states through the quantum
mechanical Gibbs-correction factor $N!$ to account for the
indistinguishability of the bosonic cells. Thus, with this
quantum-mechanical correction we obtain the number of chain-states
\begin{equation}
\Omega (N) =\frac{N^N}{N!} \; .
\label{NOS}
\end{equation}

Let us now determine from $\Omega(N)$ the entropy ${\cal S}_c$ of
the chain-states in the thermodynamic large $N$ limit. Using
Stirling's approximation, $\ln(N!)=N\ln N-N+{\cal O}(\ln N)$, we
obtain
\begin{equation}
{\cal S}_c=\ln\Omega(N)=N+{\cal O}(\ln N) \; .
\end{equation}
By virtue of (\ref{SBH4}) $N$ is however nothing else but the
BH-entropy such that finally we get
\begin{equation}
{\cal S}_c = {\cal S}_{BH}+{\cal O}(\ln {\cal S}_{BH}) \; .
\end{equation}
Thus the entropy of the proposed chain-states coincides exactly,
up to a logarithmic correction, with the semiclassical D=4
BH-entropy. Notice that in this approach there is no need to fix
the proportionality constant as is the case in many other
approaches which derive the BH-entropy.

\section{Corrections to the BH Area Law}

Recently also corrections to the BH area-law became available.
Corrections for D=4 black holes have been determined in
supersymmetric cases from string-theory (see e.g.~\cite{CWM})
while results in non-supersymmetric cases came from the Quantum
Geometry approach \cite{KM} or the Conformal Field Theory (CFT)
approach of Carlip \cite{CLC}. The general result is a logarithmic
correction, $-k\ln{\cal S}_{BH}$ with a positive constant $k>0$.
The appearance of a negative correction can be attributed to the
Holographic Principle \cite{HP} as emphasized in \cite{MHP}. For
example in the CFT approach by determining corrections to the
Cardy formula \cite{Car} one arrives at an entropy
\begin{equation}
{\cal S}_{\text{CFT}} ={\cal S}_{BH}-\frac{3}{2}\ln{\cal
S}_{BH}+\ln c+\text{const}
\end{equation}
for a class of D=4 black holes \cite{CLC}. Now $c$ the central
charge is given by
\begin{equation}
c=\frac{3A_H}{2\pi G_4}\frac{\gamma}{\kappa} \; ,
\end{equation}
where $\kappa$ is the black hole's surface gravity and $\gamma$ an
undetermined periodicity parameter. If one could assume that
$\gamma$ could be chosen such that $c$ is a constant, independent
of $A_H$, then $k=3/2$. However, it has been demonstrated in
\cite{JY} that $\gamma$ should equal $2\pi T_H$ with $T_H$ the
Hawking-temperature. Using $T_H=\frac{\kappa}{2\pi}$ one then
arrives at the value $k=1/2$.

Before addressing corrections to the BH-entropy let us emphasize
that we are working throughout this paper with a microcanonical
ensemble in equilibrium. That means we are considering the chains
as quantum microstates which share the same fixed energy. It can
be easily seen e.g.~for the Schwarzschild black hole that its
energy/mass depends like $M_{BH}\propto \sqrt{N}$ on $N$ (see
\cite{K3}). Hence fixing the energy amounts to fixing $N$. The
microscopic entropy coming from chain states ${\cal S}_c$ is then
simply defined as the logarithm of the number of chains with same
fixed energy $E$ resp.~same $N$. This number is given by
$\Omega(N)$. The corrections which we will study will arise from
using more accurate approximations to $\Omega(N)$ in the large
$N\gg 1$ regime.

We had found the expression (\ref{NOS}) for the number of states
and derived from it the chain's entropy at leading order in $N$ by
using Stirling's approximation for $N!$. This gave agreement with
the BH-entropy at this leading order. Obviously a more accurate
evaluation of the chain's entropy and hence corrections to the
BH-entropy will arise from taking a more accurate approximation
for $N!$ using the Stirling series to higher orders. This will
give a more accurate evaluation of $\Omega(N)$. For instance if we
take \cite{Arf}
\begin{equation}
N!=\sqrt{2\pi N}N^Ne^{-N}\big( 1+\frac{1}{12N}
+{\cal O}\big(\frac{1}{N^2}\big) \big)
\end{equation}
we obtain for the microcanonical chain-entropy
\begin{equation}
{\cal S}_c = \ln \Omega(N) = N-\frac{1}{2}\ln N
-\ln\sqrt{2\pi}-\frac{1}{12N}
+{\cal O}\big( \frac{1}{N^2} \big) \; .
\end{equation}
By means of the identification (\ref{SBH4}) this leads to the
corrected chain-entropy formula
\begin{equation}
{\cal S}_c = {\cal S}_{BH}-\frac{1}{2}\ln{\cal S}_{BH}
-\ln\sqrt{2\pi}-\frac{1}{12{\cal S}_{BH}}
+{\cal O}\big( \frac{1}{{\cal S}_{BH}^2} \big) \; .
\end{equation}
The chain-entropy gives therefore not only the expected leading
logarithmic correction term but also agrees quantitatively with
$k=1/2$. Moreover, the first three correction terms are negative
in accord with the aforementioned restriction from the Holographic
Principle.

\section*{Acknowledgments}

The author would like to thank Alex Kehagias and Elias Kiritsis
for related discussions. This work has been supported by the
National Science Foundation under Grant Number PHY-0099544 and by
the European Community's Human Potential Program under contract
HPRN-CT-2000-00148.

\newcommand{\zpc}[3]{{\it Z.~Phys.} {\bf C\,#1} (#2) #3}
\newcommand{\npb}[3]{{\it Nucl.~Phys.} {\bf B\,#1} (#2) #3}
\newcommand{\npps}[3]{{\it Nucl.~Phys.~Proc.~Suppl.} {\bf #1} (#2) #3}
\newcommand{\plb}[3]{{\it Phys.~Lett.} {\bf B\,#1} (#2) #3}
\newcommand{\prd}[3]{{\it Phys.~Rev.} {\bf D\,#1} (#2) #3}
\newcommand{\prb}[3]{{\it Phys.~Rev.} {\bf B\,#1} (#2) #3}
\newcommand{\pr}[3]{{\it Phys.~Rev.} {\bf #1} (#2) #3}
\newcommand{\prl}[3]{{\it Phys.~Rev.~Lett.} {\bf #1} (#2) #3}
\newcommand{\prsla}[3]{{\it Proc.~Roy.~Soc.~Lond.} {\bf A\,#1} (#2) #3}
\newcommand{\jhep}[3]{{\it JHEP} {\bf #1} (#2) #3}
\newcommand{\cqg}[3]{{\it Class.~Quant.~Grav.} {\bf #1} (#2) #3}
\newcommand{\grg}[3]{{\it Gen.~Rel.~Grav.} {\bf #1} (#2) #3}
\newcommand{\prep}[3]{{\it Phys.~Rep.} {\bf #1} (#2) #3}
\newcommand{\fp}[3]{{\it Fortschr.~Phys.} {\bf #1} (#2) #3}
\newcommand{\nc}[3]{{\it Nuovo Cimento} {\bf #1} (#2) #3}
\newcommand{\nca}[3]{{\it Nuovo Cimento} {\bf A\,#1} (#2) #3}
\newcommand{\lnc}[3]{{\it Lett.~Nuovo Cimento} {\bf #1} (#2) #3}
\newcommand{\ijmpa}[3]{{\it Int.~J.~Mod.~Phys.} {\bf A\,#1} (#2) #3}
\newcommand{\ijmpd}[3]{{\it Int.~J.~Mod.~Phys.} {\bf D\,#1} (#2) #3}
\newcommand{\rmp}[3]{{\it Rev.~Mod.~Phys.} {\bf #1} (#2) #3}
\newcommand{\ptp}[3]{{\it Prog.~Theor.~Phys.} {\bf #1} (#2) #3}
\newcommand{\sjnp}[3]{{\it Sov.~J.~Nucl.~Phys.} {\bf #1} (#2) #3}
\newcommand{\sjpn}[3]{{\it Sov.~J.~Particles\& Nuclei} {\bf #1} (#2) #3}
\newcommand{\splir}[3]{{\it Sov.~Phys.~Leb.~Inst.~Rep.} {\bf #1} (#2) #3}
\newcommand{\tmf}[3]{{\it Teor.~Mat.~Fiz.} {\bf #1} (#2) #3}
\newcommand{\jcp}[3]{{\it J.~Comp.~Phys.} {\bf #1} (#2) #3}
\newcommand{\cpc}[3]{{\it Comp.~Phys.~Commun.} {\bf #1} (#2) #3}
\newcommand{\mpla}[3]{{\it Mod.~Phys.~Lett.} {\bf A\,#1} (#2) #3}
\newcommand{\cmp}[3]{{\it Comm.~Math.~Phys.} {\bf #1} (#2) #3}
\newcommand{\jmp}[3]{{\it J.~Math.~Phys.} {\bf #1} (#2) #3}
\newcommand{\pa}[3]{{\it Physica} {\bf A\,#1} (#2) #3}
\newcommand{\nim}[3]{{\it Nucl.~Instr.~Meth.} {\bf #1} (#2) #3}
\newcommand{\el}[3]{{\it Europhysics Letters} {\bf #1} (#2) #3}
\newcommand{\aop}[3]{{\it Ann.~of Phys.} {\bf #1} (#2) #3}
\newcommand{\arnps}[3]{{\it Ann.~Rev.~Nucl.~Part.~Sci.} {\bf #1} (#2) #3}
\newcommand{\jetp}[3]{{\it JETP} {\bf #1} (#2) #3}
\newcommand{\jetpl}[3]{{\it JETP Lett.} {\bf #1} (#2) #3}
\newcommand{\acpp}[3]{{\it Acta Physica Polonica} {\bf #1} (#2) #3}
\newcommand{\sci}[3]{{\it Science} {\bf #1} (#2) #3}
\newcommand{\nat}[3]{{\it Nature} {\bf #1} (#2) #3}
\newcommand{\pram}[3]{{\it Pramana} {\bf #1} (#2) #3}
\newcommand{\hepth}[1]{{\tt hep-th/}{\tt #1}}
\newcommand{\hepph}[1]{{\tt hep-ph/}{\tt #1}}
\newcommand{\grqc}[1]{{\tt gr-qc/}{\tt #1}}
\newcommand{\astroph}[1]{{\tt astro-ph/}{\tt #1}}
\newcommand{\desy}[1]{{\it DESY-Report}{#1}}

\end{document}